\documentclass{article}

\usepackage{arxiv}

\usepackage[utf8]{inputenc} 
\usepackage[T1]{fontenc}    
\usepackage{hyperref}       
\usepackage{url}            
\usepackage{booktabs}       
\usepackage{amsfonts}       
\usepackage{nicefrac}       
\usepackage{microtype}      
\usepackage{cleveref}       
\usepackage{lipsum}         
\usepackage{graphicx}
\usepackage{natbib}
\usepackage{doi}

\usepackage{listings}
\usepackage{color}

\definecolor{dkgreen}{rgb}{0,0.6,0}
\definecolor{gray}{rgb}{0.5,0.5,0.5}
\definecolor{mauve}{rgb}{0.58,0,0.82}

\graphicspath{ {./images} }

\title{Scalable Software as a Service Architecture}

\newif\ifuniqueAffiliation
\uniqueAffiliationtrue

\ifuniqueAffiliation 
\author{ \href{https://ardy.me}{\includegraphics[scale=0.06]{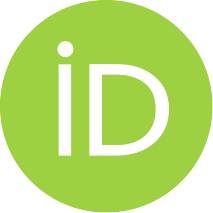}\hspace{1mm}Ardy Dedase}\thanks{Use footnote for providing further
		information about author (webpage, alternative
		address)---\emph{not} for acknowledging funding agencies.} \\	
	\texttt{ardy@u.nus.edu} \\
	\texttt{ardy.me} \\
}
\else
\usepackage{authblk}

\newbox{\orcid}\sbox{\orcid}{\includegraphics[scale=0.06]{orcid.pdf}} 
\author[1]{%
	\href{https://orcid.org/0000-0000-0000-0000}{\usebox{\orcid}\hspace{1mm}David S.~Hippocampus\thanks{\texttt{hippo@cs.cranberry-lemon.edu}}}%
}
\author[1,2]{%
	\href{https://orcid.org/0000-0000-0000-0000}{\usebox{\orcid}\hspace{1mm}Elias D.~Striatum\thanks{\texttt{stariate@ee.mount-sheikh.edu}}}%
}
\affil[1]{Department of Computer Science, Cranberry-Lemon University, Pittsburgh, PA 15213}
\affil[2]{Department of Electrical Engineering, Mount-Sheikh University, Santa Narimana, Levand}
\fi

\renewcommand{\headeright}{}

\hypersetup{
pdftitle={Scalable Software as a Service Architecture},
pdfsubject={q-bio.NC, q-bio.QM},
pdfauthor={Ardy G.~Dedase},
pdfkeywords={Software as a Service, SaaS, Architecture, Scalability, Maintainability},
}

\begin{document}
\maketitle

\begin{abstract}
This paper explores the architecture of Software as a Service (SaaS) platforms, emphasizing scalability and maintainability. SaaS, a flexible software distribution model suitable for individuals and organizations, has become prevalent with the advent of Cloud services. This paper aims to provide a high-level design reference for establishing a scalable and maintainable SaaS architecture.
\end{abstract}


\section{Introduction}

\subsection{Why Software as a Service?}
Software as a Service (SaaS) represents a versatile software distribution model suitable for individuals or solo entrepreneurs and organizations alike. 
Cloud services have empowered the operation of independent SaaS platforms, enabling businesses to build freemium models successfully. 
However, the simplicity of SaaS system design can lead to challenges without a well-thought-out architecture, resulting in monolithic applications with redundancy.

\subsection{Requirements of a SaaS Platform}
To ensure the success of a SaaS platform, certain requirements must be addressed:
\begin{itemize}
    \item Access control for users and user groups
    \item Subscription tier management
    \item Hosting internal admin tools
    \item Extensible features for easy addition of new products
    \item Design goals for isolated services, reducing the blast radius of changes, and running web applications in isolation.
\end{itemize}

\section{The Solution: Isolate and Reuse}
\label{sec:headings}

\subsection{Isolate}

Isolated web applications are formed by grouping related features together to form one web application or group of web applications that represents a product. For specificity, I would call these as \textit{Product Web Apps} which will be covered more in detail.

\subsubsection{Group related features in a web application}

For example, all \textit{analytics reporting features} can be grouped together as an independent web application that will be built and maintained by one dedicated team that has domain knowledge in building analytics reporting products. In large companies, every product web app has its own dedicated team building and running it.

\subsubsection{Internal vs public (optional)}
Your product web apps can also be grouped into two: internal and public. Having a clear separation between internal and public will allow you to have a dedicated set of secured administration tools routed within your own private network.

Proxying these services internally and publicly is handled by the Routing Service which will be covered in one of the sections below.

\subsection{Reuse}
These are the common functionalities that are shared across the feature web apps with their corresponding services. Each feature web app will need to utilise each of the functionality listed below.

\begin{tabular}{ |p{6cm}|p{6cm}|  }
	\hline
	\textbf{Functionality} & \textbf{Service}\\
	\hline
	Routing. Proxying to requested feature web applications. & Routing service\\
	\hline
	Content management & Content repository\\
	\hline
	Authentication & Authentication service\\
	\hline
	Access management & Role-based access control (RBAC) service\\	
	\hline
\end{tabular}

\section{Components of a Scalable SaaS Platform}
These are the logical components that make up our SaaS.

\subsection{Routing Service}
\begin{figure}[h]
	\centering
	\includegraphics[scale=0.35]{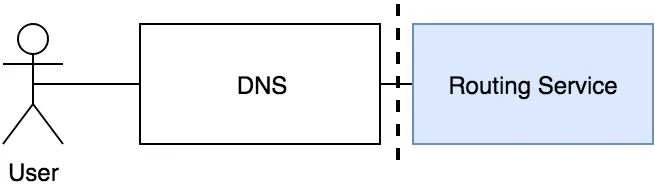}
	\caption{Routing service}
	\label{fig:router}
\end{figure}

The Routing Service directs user requests (see figure \ref{fig:router}) to the appropriate page or feature web app. 

\begin{figure}[h]
	\centering
	\includegraphics[scale=0.5]{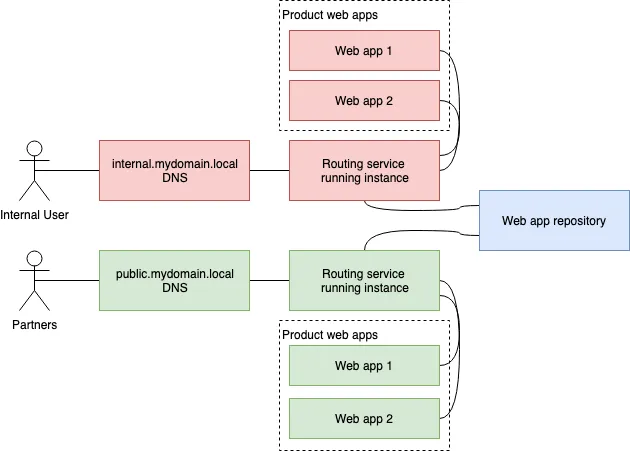}
	\caption{Routing service high-level design}
	\label{fig:router-hld}
\end{figure}

It manages mappings between paths, web application URLs, and required permissions, with the ability to update them dynamically. The service also handles the exposure of product web apps, both internally and publicly (see figure \ref{fig:router-hld}). 

Implementation Tip: Options such as Node.js's http-proxy-middleware or Nginx's Reverse proxy can be considered.

\subsection{Product Web App}

\begin{figure}[ht]
	\centering
	\includegraphics[scale=0.5]{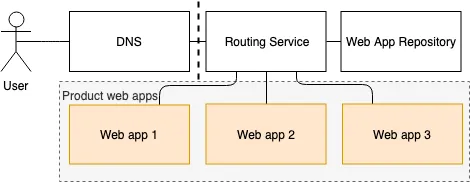}
	\caption{Product web apps high-level design}
	\label{fig:product-web-apps}
\end{figure}

Every SaaS product or feature can have its own web app that is exposed through its own URL, and these apps are grouped into internal and public categories. 
Each product web app comprises a group of related features, owned and maintained by dedicated teams (See figure \ref{fig:product-web-apps}).
The use of design systems \cite{vendramini2021towards} and reusable micro frontends \cite{pavlenko2020micro} ensures consistent user experience across products.

Implementation Tip: Product web apps can use standard web application stacks like Express, Node.js, and React, enhanced by design systems like Mosaic or Open Components.

\subsection{Web App Repository}

\begin{figure}[ht]
	\centering
	\includegraphics[scale=0.5]{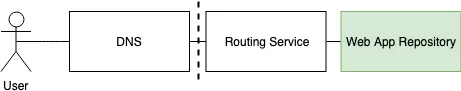}
	\caption{Web App Repository high-level design}
	\label{fig:web-app-repo}
\end{figure}

This service maintains a record of web applications owned by multiple teams. The Routing Service queries the Web App Repository to retrieve the corresponding web app URL for a requested path. Permissions information is obtained from the RBAC service to validate user access (See figure \ref{fig:web-app-repo}).

Below is an example Web app repository table to illustrate its functionality.

\begin{tabular}{ |p{4cm}|p{4cm}|p{4cm}|p{4cm}|  }
	\hline
	\textbf{Path} & \textbf{Product Web App URL} & \textbf{Owning Team} & \textbf{Description}\\
	\hline
	/my-product-1 & https://product-webapp-1.mydomain.local & Team 1 & Allows internal users to check the revenue data\\
	\hline
	/my-product-2 & https://product-webapp-1.mydomain.local & Team 2 & Allows users to set up reporting dashboards\\	
	\hline
\end{tabular}

Implementation Tip: The Web App Repository can be implemented as a RESTful service with a web admin UI or as an infrastructure as code configuration file within the Routing Service.

\subsection{Authentication}

\begin{figure}[ht]
	\centering
	\includegraphics[scale=0.5]{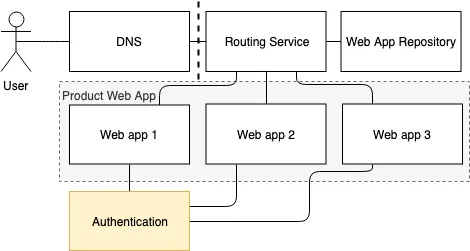}
	\caption{Authentication invoked by the Product Web Apps}
	\label{fig:auth-webapp}
\end{figure}

\begin{figure}[ht]
	\centering
	\includegraphics[scale=0.5]{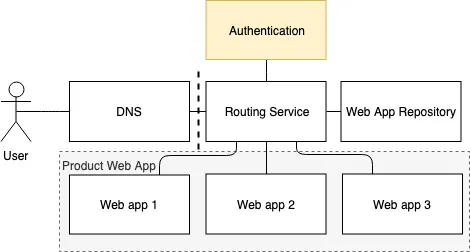}
	\caption{Authentication invoked by the Routing Service}
	\label{fig:auth-router}
\end{figure}

Authentication is handled by a dedicated service, allowing seamless passage of JSON web tokens and user information across services. 
The flexibility exists for authentication to be invoked either in the product web app \ref{fig:auth-webapp} or the Routing Service\ref{fig:auth-router}, with different trade-offs.

Implementation Tip: Services such as AWS Cognito, Auth0, or any JWT authentication service can be employed.

\subsection{Role-based Access Control (RBAC)}

After authentication, RBAC \cite{sandhu1998role} determines whether a user has permission to access specific pages or perform certain actions. 
This service is crucial for managing user permissions and ensuring controlled access.

We need to be able to answer questions like the following:

\textbf{Is the logged-in user allowed to access this page or information that they are trying to access?}

\textit{Possible answers are:}

\begin{itemize}
	\item Yes, the user Alice belongs to Organisation A which has permission to access Content A.
	\item No, the user Bob belongs to Organisation B which has no permission to access Content B.
\end{itemize}

\textbf{Is the logged-in user allowed to perform this action?}

\textit{Possible answers are:}

\begin{itemize}
	\item Yes, the user Alice has Admin Role which has permission to remove a registered user.
	\item No, the user Bob has Member Role which has read-only permission to all content.
\end{itemize}

RBAC helps us answer the questions above, hence making it possible for our web apps to permit or revoke users in doing specific actions (see figure \ref{fig:rbac}).

\begin{figure}[ht]
	\centering
	\includegraphics[scale=0.5]{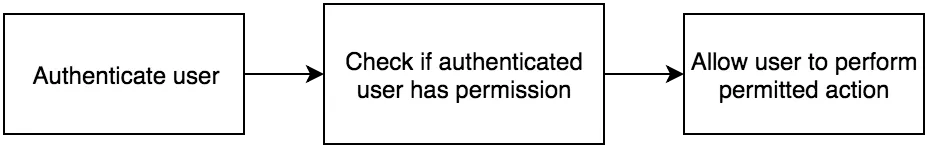}
	\caption{Role-based Access Control flow diagram}
	\label{fig:rbac-flow}
\end{figure}

Figure \ref{fig:rbac-flow} shows the flow of events that occur when a user is authenticated and has its permissions checked with RBAC.

Pseudo-code below illustrates how RBAC service can be used in specific use cases.

Check if user has permission to view page:

\begin{lstlisting}
if (userHasPermissionToViewPage(userId)) {
	showPage();
} else {
	showNoPermissionError();
}
\end{lstlisting}

Check if user has permission to perform an action:

\begin{lstlisting}
if (userHasPermissionToPerformAction(userId)) {
	showButton();
} else {
	doNothing();
}	
\end{lstlisting}

\begin{figure}[ht]
	\centering
	\includegraphics[scale=0.5]{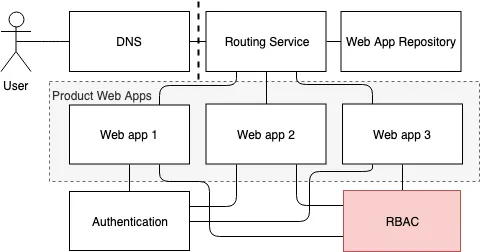}
	\caption{Role-based Access Control high-level design}
	\label{fig:rbac}
\end{figure}

Implementation Tip: Consideration for implementation choices similar to the Web App Repository, using Java, Go, or C\# with an RDBMS and a caching system.

\section{High-Level Design}

\begin{figure}
	\centering
	\includegraphics[scale=0.5]{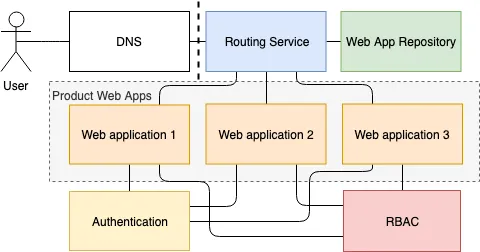}
	\caption{High-level design}
	\label{fig:high-level-design}
\end{figure}

The high-level design in Figure \ref{fig:high-level-design} emphasizes clear ownership, reuse of common functionalities, and separation of concerns. The components work in harmony to create a scalable and maintainable SaaS platform.

\section{High-Level Sequence Diagram}

Figures \ref{fig:sequence-diagram-1} and \ref{fig:sequence-diagram-2} below show our options of how a user gets routed to the requested page by the routing service. 
These events happen after the Product Web App details are retrieved from the Web App Repository.

\begin{figure}[ht]
	\centering
	\includegraphics[scale=0.5]{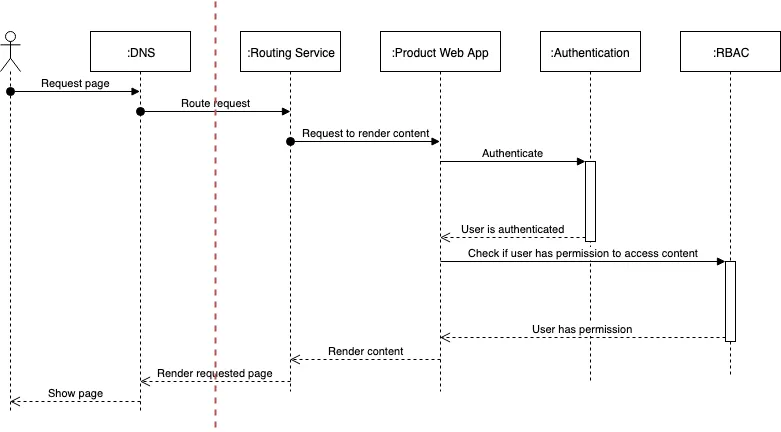}
	\caption{Invoke Authentication in the Product Web App}
	\label{fig:sequence-diagram-1}
\end{figure}

\begin{figure}[ht]
	\centering
	\includegraphics[scale=0.5]{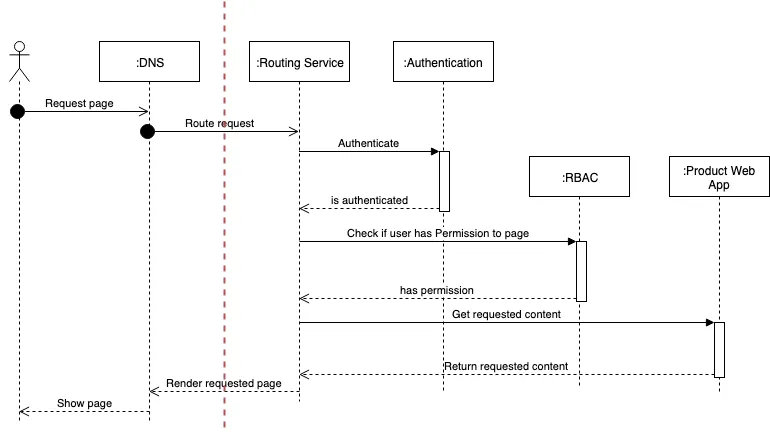}
	\caption{Invoke Authentication in the Routing Service}
	\label{fig:sequence-diagram-2}
\end{figure}

\section{Web App Repository, RBAC and Authentication page as a Web App}

\begin{figure}
	\centering
	\includegraphics[scale=0.5]{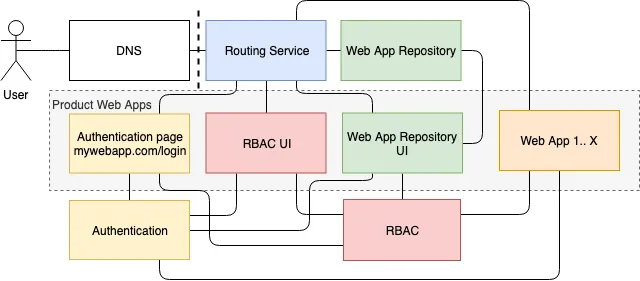}
	\caption{Some web apps are used as UIs to other components}
	\label{fig:web-apps}
\end{figure}

Web App Repository, RBAC, and Authentication need their own Web User Interface to manage data or at a minimum some way to receive user input (See figure \ref{fig:web-apps}).

Web App Repository UI allows us to manage Paths and Web App URLs with their required permissions.

RBAC UI is used to manage users, organisations, roles and permissions.

Authentication service’s frontend UI which is the login page is deployed as a Web App.

\section{Conclusion}

\subsection{Treat technical documentation as part of the product}
Enabling teams that own the Product Web Apps through good documentation whilst taking advantage of the tools that are already available is crucial. Good documentation minimises the need for your teams to go back and forth asking questions about your SaaS platform.

\subsection{Enable your fellow engineers}
On top of the web UI management tools for RBAC and Web App Repository, you can develop a NodeJS / React template that includes the clients to RBAC and Authentication out of the box, to help minimise the work needed to add a Web App to your SaaS Platform. This makes it convenient for your engineers or yourself to create a new Web App out of this template without having to spend much effort integrating it to the platform.

\subsection{Engineering design reviews}
Badly written code can cost days to weeks of developer’s time, but bad architectural decisions can cost months to years in comparison. Engineering design reviews are probably more feasible for larger companies; however if you’re a developer on your own, it doesn’t hurt to get feedback from other engineers you know in your community. Spending a reasonable amount of time writing the engineering design document upfront and getting feedback during engineering design reviews will save you more time in the future. Validate software design assumptions and get other engineers to poke as many holes as possible to your intended software architecture (no matter how tempting it might be to take some shortcuts!).

\bibliographystyle{unsrtnat}
\bibliography{references}  






\end{document}